\documentclass[11pt,twoside]{article}
\usepackage{./asp2014}
\usepackage{color}

\hypersetup{
 colorlinks= true, 
 urlcolor  = black, 
 linkcolor = red, 
 citecolor = blue 
} 

\aspSuppressVolSlug
\resetcounters

\begin{document}

\title{Young Radio AGN in the ngVLA Era}
\author{Pallavi Patil$^{1,2}$, Kristina Nyland$^1$, Jeremy J. Harwood$^3$, Amy Kimball$^4$, and Dipanjan Mukherjee$^5$}

\affil{$^1$National Radio Astronomy Observatory, Charlottesville, VA, 22903; \email{knyland@nrao.edu}; \email{ppatil@nrao.edu}}
\affil{$^2$University of Virginia, Department of Astronomy, Charlottesville, VA 22903, USA; \email{pp3uq@virginia.edu}}
\affil{$^3$Centre for Astrophysics Research, School of Physics, Astronomy and Mathematics, University of Hertfordshire, College Lane, Hatfield AL10 9AB, UK; \email{jeremy.harwood@physics.org}}
\affil{$^4$National Radio Astronomy Observatory, Socorro, NM, 87801; \email{akimball@nrao.edu}}
\affil{$^5$Dipartimento di Fisica Generale, Universita degli Studi di Torino, Via Pietro Giuria 1, 10125 Torino, Italy; \email{dipanjan.mukherjee@unito.it}}

\paperauthor{Pallavi Patil}{pp3uq@virginia.edu}{University of Virginia}{National Radio Astronomy Observatory}{}{Charlottesville}{VA}{22903}{USA}

\begin{abstract}
Most massive galaxies are now thought to go through an Active Galactic Nucleus (AGN) phase one or more times. Yet, the cause of triggering and the variations in the intrinsic and observed properties of AGN population are still poorly understood. 
Young, compact radio sources associated with accreting supermassive black holes (SMBHs) represent an important phase in the life cycles of jetted AGN for understanding AGN triggering and duty cycles. The superb sensitivity and resolution of the ngVLA, coupled with its broad frequency coverage, will provide exciting new insights into our understanding of the life cycles of radio AGN and their impact on galaxy evolution. 
The high spatial resolution of the ngVLA will enable resolved mapping of young radio AGN on sub-kiloparsec scales 
over a wide range of redshifts.  With broad continuum coverage from 1 to 116 GHz, the ngVLA will excel at estimating ages of sources as old as $30-40$~Myr at $z \sim 1$. 
In combination with lower-frequency ($\nu < 1$~GHz) instruments such as ngLOBO and the Square Kilometer Array, the ngVLA will robustly characterize the spectral energy distributions of young radio AGN.



\end{abstract}

\section{Introduction}

Young, compact radio sources associated with accreting massive black holes (MBHs) represent a key phase in the life cycles of jetted active galactic nuclei (AGN). 
These objects have recently (re)ignited their central engines within the past 10$^2$ to 10$^4$ years, resulting in sub-galactic jet extents of $\sim$10~pc to a few kpc. 
One of the unique characteristics of young radio AGN is their convex radio spectral energy distributions (SEDs) 
and characteristic spectral turnovers that peak at frequencies in the range of hundreds of MHz to a few GHz.  
Well-studied classes of young radio AGN (\citealt{odea+98, orienti+16} and references therein) include Compact Steep-Spectrum (CSS) sources, Gigahertz-peaked Spectrum (GPS) sources, and High-frequency Peakers (HFP).  
These 
young radio AGN 
may eventually evolve to become large and powerful FR I/FR II galaxies \citep{fanaroff+74,snellen+00,kunert+10}. 

An open question regarding young radio AGN is the level of their energetic impact on the interstellar medium (ISM) of the host galaxy. Since their radio jets are fully contained within the host galaxy, spatially-resolved observations are necessary. 
A growing number of studies have reported detections of multiphase outflow signatures associated with young radio AGN engaged in jet-ISM feedback (e.g.\citealt{chandola+11, holt+11, morganti+13}), but the importance of this feedback in the context of galaxy evolution remains unclear. The ngVLA will address this issue through both continuum and spectral line observations. Broad-band continuum measurements of the turnover frequency of the radio SEDs of young radio sources will also directly constrain the density distribution of the ISM for sources at low redshift (Section~\ref{sec:turno}; \citealt{bicknell+97, bicknell+17, jeyakumar+16}). Follow-up spectral line observations probing the conditions of the ISM in the vicinity of the jets of the young radio AGN will further constrain the kinematics of the gas, thus probing the energetic impact of the jet-driven feedback.

Young radio AGN also hold vital clues about radio triggering and duty cycles, which are still poorly understood \citep{tadhunter+16}. 
An 
improved understanding of MBH duty cycles and MBH accretion mechanisms will help explain the impact of MBH growth on 
galaxy 
growth and evolution \citep{shulevski+15}. Molecular gas line or HI absorption diagnostics can establish a link between fueling and different triggering mechanisms \citep{maccagni+14}. A significant amount of dense gas is seen in young radio AGN \citep{odea+98}, and several studies have found that H{\tt I} is detected more frequently in GPS and CSS sources compared to other types of radio AGN (\citealt{gereb+14}; \citealt{odea+98} and references therein). Therefore, the investigation of these newborn radio AGN can give us a direct view of many transient processes and better inform simulations of galaxy formation and evolution.  

\section{Anticipated Results}
\subsection{Resolving Young Radio AGN with the ngVLA}

\begin{figure*}
\centering
\includegraphics[clip=true, trim=0.05cm 5cm 0.05cm 7cm, height=0.37\textwidth]{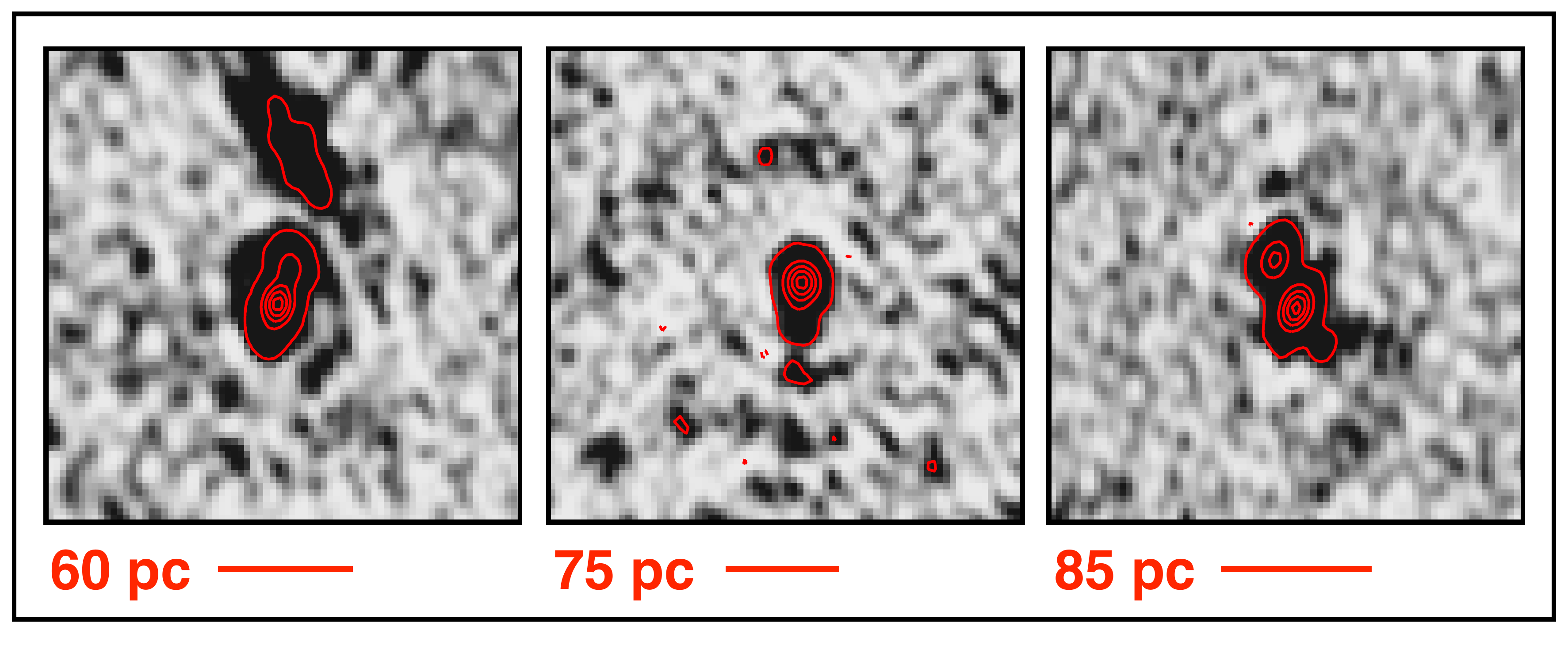}
\caption{Example Very Long Baseline Array continuum images (\citealt{nyland+18}; Patil et al.~in preparation) of young, compact radio AGN with extents of tens of pc.  These sources are drawn from the sample presented in \citet{lonsdale+15}, which selects heavily obscured, powerful AGN embedded in luminous starbursting galaxies at intermediate redshifts with compact radio emission based on their properties from the Wide-Field Infrared Survey Explorer \citep{wright+10} and the NRAO--VLA Sky Survey \citep{condon+98}.  The high collecting area and angular resolution of the ngVLA will facilitate deep, spatially-resolved surveys of compact radio jets with comparable extents in a wide range of systems, providing new insights into the energetic impact of young radio jets under a variety of host galaxy conditions and properties.}
\label{fig:pallavis_sources}
\end{figure*}

In Figure~\ref{fig:pallavis_sources}, we show examples of young radio AGN candidates with jet extents of 60 to 85~pc. The observations are taken with the Very Long Baseline Array at C-band (Patil et al., in preparation). These radio AGN were drawn from the sample of extremely infrared luminous sources identified in \citet{lonsdale+15} that are believed to be in a unique evolutionary stage just after the (re)ignition of the radio AGN but while the host galaxy is going through a starburst phase due to 
a recent merger. The angular resolution, frequency range and sensitivity offered by the ngVLA are optimal for efficiently resolving the structures of young radio AGN, such as the sources highlighted in Figure~\ref{fig:pallavis_sources}, and constraining their ages through broadband continuum observations (Section~\ref{sec:ages}). 

Simulations of relativistic jets interacting with ISM \citep{mukherjee+16,bicknell+17,mukherjee+18} have shown that jets drive shocks in the ISM via an energy bubble. Furthermore, \citet{mukherjee+18} have found that jet powers, inclinations, and densities affect AGN-driven outflows, thus motivating the need for high-resolution radio imaging to obtain spatially-resolved structural, spectral, and polarimetric information on the radio jets. 
In addition, direct measurements of flux densities and source sizes provide estimates of the energy density and pressure in the synchrotron plasma, which can then constrain ISM densities and the rate of the advancing shock front into the ambient medium. Limits on Faraday rotation measures obtained through polarimetry help constrain models of the nuclear environment. 

The planned longest baselines of the ngVLA of up to $\sim$1000~km 
are required to provide the 
subarcsecond resolution needed to study the population of young radio AGN. 
The ngVLA will be able to easily detect radio AGN with sub-kiloparsec structures over a wide redshift range \citep{nyland+18}. 
The significant increase in 
point source sensitivity compared to Jansky Very Large Array (JVLA) will enable the efficient detection of radio emission in young radio AGN.  
An interesting point of comparison is the Legacy eMERLIN Multi-band Imaging of Nearby Galaxies survey
(LeMMINGs; \citealt{baldi+18}) that aims to probe the radio properties of nearby AGN at sub-arcsecond spatial resolution.  In the same on-source integration time used in the LeMMINGs survey (48 minutes), the ngVLA would be able to reach a depth of about 100 times deeper ($\sim$0.7~$\mu$Jy~beam$^{-1}$) in its Band~1 ($1.2 - 3.5$). 

\subsection{Radio Spectral Ages}
\label{sec:ages}

\begin{figure*}
\centering
\includegraphics[clip=true, trim=0.5cm 5.25cm 2.7cm 6cm, height=0.475\textwidth]{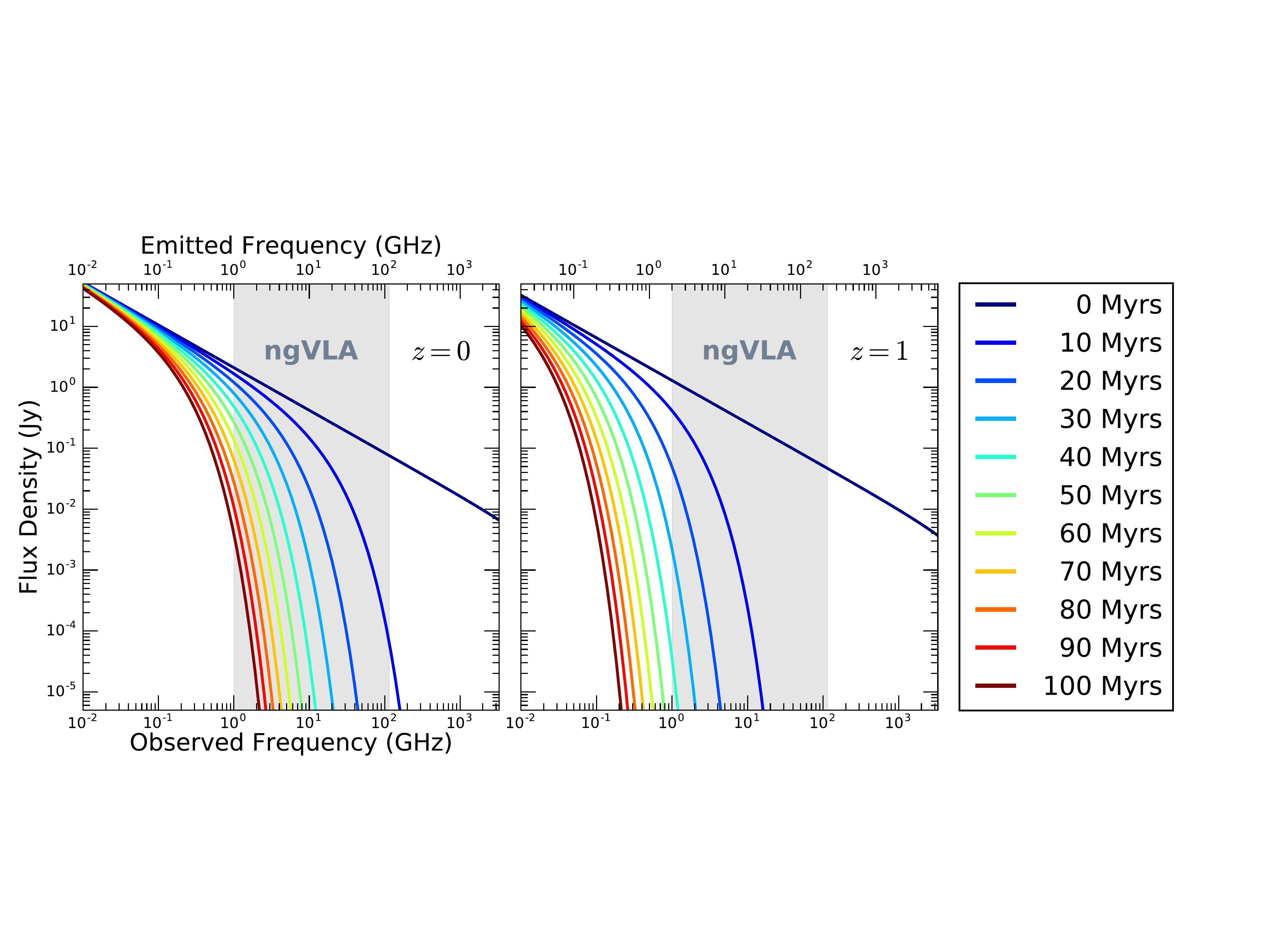}
\caption{Example of JP model \citep{jaffe+73} spectral ages calculated using the BRATS software \citep{harwood+13} demonstrating the need for ngVLA observations spanning a wide range of frequencies. The left and center panels correspond to redshifts of 0 and 1, respectively.  The flux density values shown on the $y$-axis have been arbitrarily scaled.  Because of its advantages of the wide frequency range and angular resolution compared to the SKA, the ngVLA will uniquely excel in studies of low-redshift radio AGN that are young, or higher-redshift AGN that are embedded in dense environments.}
\label{fig:JP}
\end{figure*}

For a fixed magnetic field, if the initial energy distribution of a population of electrons follows a power law, given by $N(E) = N_0 E^{-\delta}$, the energy losses scale as $\tau = \frac{E}{dE/dt} \propto 1/E \propto 1/\nu^{2}$.  This results in a preferential cooling of higher energy electrons. Assuming no other mode of particle acceleration, the resulting spectrum becomes increasingly curved over time. This behavior can be used to determine the characteristic age of a radio source (e.g. \citealt{myers+85, harwood+13}).																		

In Figure~\ref{fig:JP}, we show an example 
of spectral ages calculated using the BRATS software\footnote{http://www.askanastronomer.co.uk/brats/} \citep{harwood+13} for an arbitrary jetted radio AGN at $z = 0$ and $z = 1$. As illustrated by the spectral age curves in the figure, the ngVLA will excel in studies of radio AGN spanning a broad range of ages at low redshift, as well as young or embedded radio AGN at higher redshifts. The results from the Australia Telescope 20~GHz (AT20G) survey \citep{murphy+10} indicate that continuum measurements in the tens of GHz range are needed to model the radio spectral energy distributions adequately \citep{sadler+06,sadler+08}. These observations are particularly crucial for modeling the ages of young, low-redshift sources less than 10~Myrs old.  Lower frequency radio continuum data in the MHz range are important for constraining the ages of high-$z$ sources; however, the inclusion of the lowest-frequency ngVLA bands down to $\sim$1~GHz would provide sufficient frequency coverage for measuring the ages of sources as old as 30-40~Myrs at $z \sim 1$. 

\subsection{Characterizing Spectral Turnover}\label{sec:turno}

CSS, GPS, and HFP sources show an inverse correlation between the spectral peak frequency and their linear sizes \citep{odea+98, orienti+14}. 
The spectra may be peaked due to the absorption by either a relativistic plasma or an external screen of ionized gas. Simulations by \citealt{bicknell+17} have shown that the location of each source on the turnover-linear size relation is dependent on the warm ISM density, suggestive of an evolutionary link between all three classes of young radio AGN. However, the details of the absorption model remain unclear
\citep{callingham+15}. Therefore, high-resolution observations sampling both sides of the peak/turnover frequency are essential for identifying the absorption process. 

\begin{figure*}
\centering
\includegraphics[clip = true, trim = 0cm 0.cm 0.05cm 0.05cm,  width=\textwidth]{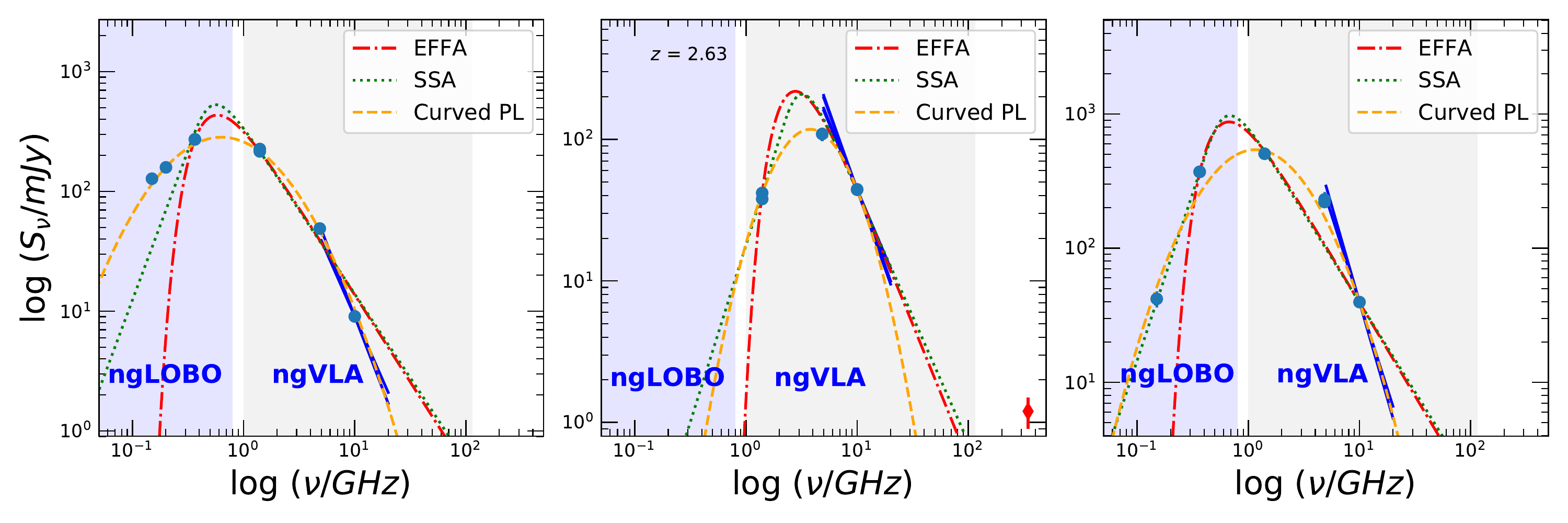}
\caption{An example of radio SED fitting in GPS/HFP sources. These sources are taken from a parent sample of heavily obscured and hyper-luminous quasars with compact radio emission \citep{lonsdale+15}. The data points at 10~GHz are from a high-resolution JVLA imaging study of young radio AGN (Patil et al., in preparation). The blue line segments show the in-band spectral indices of the JVLA data. The remaining data points are from published radio surveys. The source in the second panel has a known redshift, and the red diamond is the ALMA Band 7 observation taken in cycle 0. The curves are the best-fit solutions of different radio SED models.  A red dashed dot line is for the external free-free absorption model (EFFA), a dotted green line corresponds to the synchrotron self-absorption model (SSA), and the yellow dashed line is for the generic curved power law model. The ngVLA and ngLOBO will map the radio SED in these sources and will be able to distinguish between different theoretical models.}
\label{fig:rsed}
\end{figure*}

As shown in Figure~\ref{fig:rsed}, the inclusion of low-frequency observations around 1~GHz or lower would enhance studies of the turnover in the optically-thick radio SEDs of young radio AGN.  This analysis would provide new insights into the dominant physical mechanism responsible for the turnover (e.g. synchrotron self-absorption vs. free-free absorption; \citealt{ken+66}) in young radio AGN over a larger parameter space.  The Next Generation LOw Band Observatory (ngLOBO; \citealt{taylor+17}), which is a proposed enhancement to the basic ngVLA reference design, would extend the ngVLA's frequency range below 1~GHz. NgLOBO would enable more robust radio SED and spectral aging model studies with the ngVLA, particularly for slightly older ($>$ 10~Myr) and more distant ($z > 1$) sources.

\section{Uniqueness to ngVLA Capabilities}
The broad frequency range (1--116~GHz) of the ngVLA will facilitate surveys of the radio spectral ages of {\it young} AGN residing in gas-rich galaxies at low and intermediate redshifts.  We emphasize that unlike other next-generation radio facilities that will operate primarily at lower frequencies (e.g. the SKA), 
the ngVLA will be uniquely well-suited for measuring the ages of young radio AGN potentially engaged in jet-ISM feedback.

\section{Multiwavelength Synergy}

Young radio AGN are an important class of objects for studying the interplay between galaxies and their central engines.  
Upcoming or current facilities in the radio regime, as well as at other wavelengths, will be essential for probing the morphologies, evolutionary stages, and energetics of young radio AGN, and will ultimately provide new insights into radio jet physics and its impact on galaxy growth and evolution. 

\subsection{SKA}
The ngVLA and Square Kilometre Array (SKA) will both allow broadband studies of radio SEDs over their respective wide frequency ranges, thus providing the temporal information from spectral aging models necessary for interpreting the evolutionary stage of the source. The SKA will be able to detect emission associated with fading radio lobes, tracing the regime of radio AGN with older spectral ages (``AGN archaeology;'' \citealt{morganti+17}). In Figure~\ref{fig:JP}, we showed that the 1--116 GHz observing range of the ngVLA is advantageous for studies of low-redshift radio AGN, or intermediate redshift sources that are very young. Such young sources are often embedded in dense environments, and are known to drive multi-phase gas outflows \citep[e.g.][]{holt+08,gereb+15}. Broadband radio continuum surveys with both the ngVLA and SKA will therefore be needed to construct a complete picture of the life cycles of radio AGN and their connection to galaxy evolution. 

\subsection{ALMA}
The frequency range of the ngVLA samples non-thermal synchrotron emission associated with radio AGN and recent star formation as well as thermal emission from current star-forming regions. The higher frequency bands of the Atacama Large Millimeter/Submillimeter Array (ALMA) are sensitive to the dust continuum, which can be used to help determine robust star formation rates.  In terms of spectral line emission, the ngVLA will be able to detect the bulk of the molecular ISM through the low-J transitions of the CO molecule, while ALMA will be sensitive to higher-J CO transitions as well as a wide variety of additional molecular species and transitions that will characterize the ISM conditions.  Thus, the combination of the ngVLA and ALMA will be important for constraining the impact of young radio AGN on the ISM and star formation on galactic scales.

\subsection{Optical and Infrared Telescopes:}
Next-generation optical and infrared telescopes such as the \textit{James Webb} Space Telescope \textit{(JWST)}, Thirty Meter Telescope (TMT), Giant Magellan Telescope (GMT), and the Wide-Field Infrared Survey Telescope \textit{(WFIRST)} will provide a wealth of complementary information about the physical conditions of young radio AGN and the properties of their hosts. 
This includes resolving host morphologies, studying the kinematics of the warm ionized gas via spectroscopy, delineating star-formation and AGN via mid-infrared imaging and SED analyses, and tracing AGN-driven outflows and their excitation mechanisms. 




\bibliography{ngVLA_young_radio_AGNs_v1}  

\end{document}